\documentclass[final,1p,times]{elsarticle}

\usepackage{amssymb}
\usepackage{gensymb}
\usepackage{upgreek}
\usepackage{graphicx}
\usepackage{caption}
\usepackage{subcaption}

\usepackage{dblfloatfix}
\usepackage{fixltx2e}

\journal{Journal of Colloid and Interface Science}

\begin{document}

\begin{frontmatter}



\title{Statistical Characterization of Self-Assembled Colloidal Crystals by Single-Step Vertical Deposition}


\author{Tero J. Isotalo\corref{cor1}}
\ead{tero.isotalo@jyu.fi}
\author{Yao-Lan Tian}
\author{Mikko P. Konttinen}
\author{Ilari J. Maasilta}

\address{Nanoscience Center, Department of Physics, P.O. Box 35, FI-40014 University of Jyv\"{a}skyl\"{a}, Finland}

\begin{abstract}
We have statistically characterized the self-assembly of multi-layer polystyrene colloidal crystals, using the technique of vertical deposition, with  
parameters chosen  to produce thick layers of self-assembled crystals in one deposition step.  In addition, using a lithographically directed self-assembly method, we have shown that the size of multi-layer, continuous crack-free domains in lithographically defined areas can be many times larger than in the surrounding areas.  In a single deposition step, we have produced continuous colloidal crystal films of 260 nm diameter polystyrene spheres approximately 30 -- 40 layers thick, with a controllable lateral size of 80 -- 100 $\upmu$m without lithography, and as high as 250 $\upmu$m with the lithographic template.
\end{abstract}


\end{frontmatter}
\newpage




\section{Introduction}
\label{intro}
Colloidal crystallization of mono-disperse spherical sub-micron particles has been used successfully  in the fabrication of three-dimensional (3D) photonic crystals \cite{Jiang1999,vlasov} for a wide variety of applications \cite{Joannopoulos1997}.  These artificial opals, as they are commonly known, can also exhibit an acoustic or a phononic band gap, whose frequency depends on the particle diameter.  While the properties of 3D phononic crystals have recently been studied by various groups in the millimeter and micrometer scale at acoustic frequencies \cite{Liu2000,Yang2002,Yang2004}, much less work has been performed on 3D hypersonic phononic crystals with 100 nm scale dimensions, which produce phononic band gaps in the GHz range \cite{naturemat}. This frequency range is even relevant for thermal properties, as the dominant thermal phonons at low temperatures below 1K have frequencies in that same range \cite{zen}. 
One possibility for GHz 3D phononic crystal fabrication is therefore the use of polystyrene (PS) nano-spheres of $\sim $ 100 nm size range, as they are commercially available as colloidal suspensions. As demonstrated before \cite{Jiang1999,gu,fustin,Shimmin2006}, these particles can be self-assembled into crystals from suspensions,  allowing a convenient bottom-up assembly of a GHz phononic crystal with possible applications in thermal management or phononic waveguiding \cite{Pennec2010}.
 
Common methods for self-assembly of colloidal particles include gravitational sedimentation, capillary assembly and vertical deposition \cite{review}.  Typically, closed-packed lattices (face-centered cubic, hexagonal closed packed, or a random combination of the two) are produced in each of these methods, with face-centered cubic being more common.  Here, we use the vertical deposition method for producing large area, thick 3D closed packed lattices using colloidal suspensions of PS nanospheres. This technique in its simplest form has been documented in the literature \cite{Jiang1999,gu,fustin,Shimmin2006} quite extensively. Here, we concentrate on increasing the lateral single crystal domain size for the crystals and characterize the domain size results statistically, as a function of dipping speed and nanosphere concentration. 
  We also combine the verical deposition with lithographically patterned substrates, not to control the crystal structure itself as has been suggested before \cite{vanBlaaderen,Xia2003,Khanh2009,Zhang2010}, but to increase the size of single crystalline domains and to reduce cracking. In contrast, most other lithography directed methods \cite{Xia2003,Khanh2009,Zhang2010} are typically mono-layer techniques.  These have the drawbacks of multiple process steps and lengthy fabrication time, if thicker crystals are required.  


\section{Material and methods}
\label{MatMeth}
\begin{figure}[h]
\centering
\includegraphics[width=0.5\columnwidth]{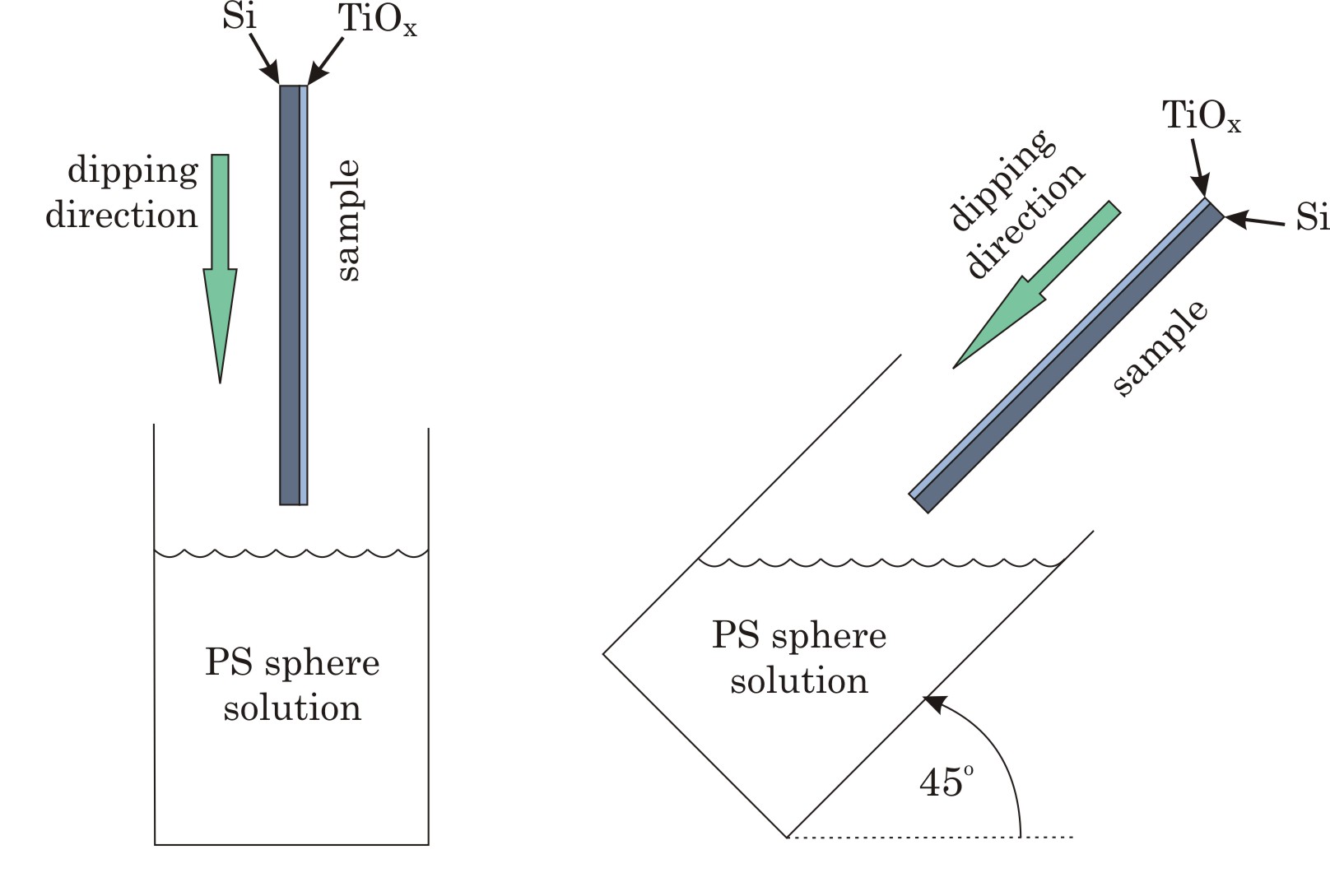}
\caption{Schematics of the dipping setups are shown. TiO$_x$ coated substrates were dipped vertically and also at a 45$\degree$ angle.}
\label{fig:DipDiagram}
\end{figure}

Samples were coated with a solution of PS nanospheres using vertical deposition \cite{Jiang1999} and the resulting crystals were characterized by scanning electron microscopy (SEM).  All substrates used in the experiments were either oxidized or nitridized silicon chips with dimensions of approximately 8 x 16 mm, and coated with a layer of titanium oxide (TiO$_x$) approximately 10 nm thick on the surface.  A titanium coating was first applied by electron-beam evaporation in an UHV chamber with a pressure 10$^{-8}$ mbar, and the resulting film was then oxidized in a vacuum chamber at 100 mbar of oxygen pressure for 2 minutes.  The TiO$_x$ provided a hydrophilic surface for improved wetting by the nanosphere solution \cite{Karuppuchamy2005}.
A schematic of the dipping setup is shown in Figure \ref{fig:DipDiagram}.
Samples without lithographically patterned surfaces were submerged into and withdrawn from solutions with concentrations of 0.02\%, 0.2\%, 2\%, 5\% and 10\%.  The original solution of 10\% concentration, purchased from Duke Scientific, was diluted with de-ionized water to produce the lower concentrations.  Each concentration was tested at withdrawal speeds from 0.01 mm/min up to 0.04 mm/min.  It was determined that the speed of withdrawal from the solution was the dominant factor in self-assembly and thus, a higher speed was used for submersion in order to reduce total dipping time.  In addition to vertical dipping, some experiments were performed at a 45$\degree$ angle for select combinations of dipping speed and solution concentration.
\begin{figure}[h]
\centering
\includegraphics[width=1.0\textwidth]{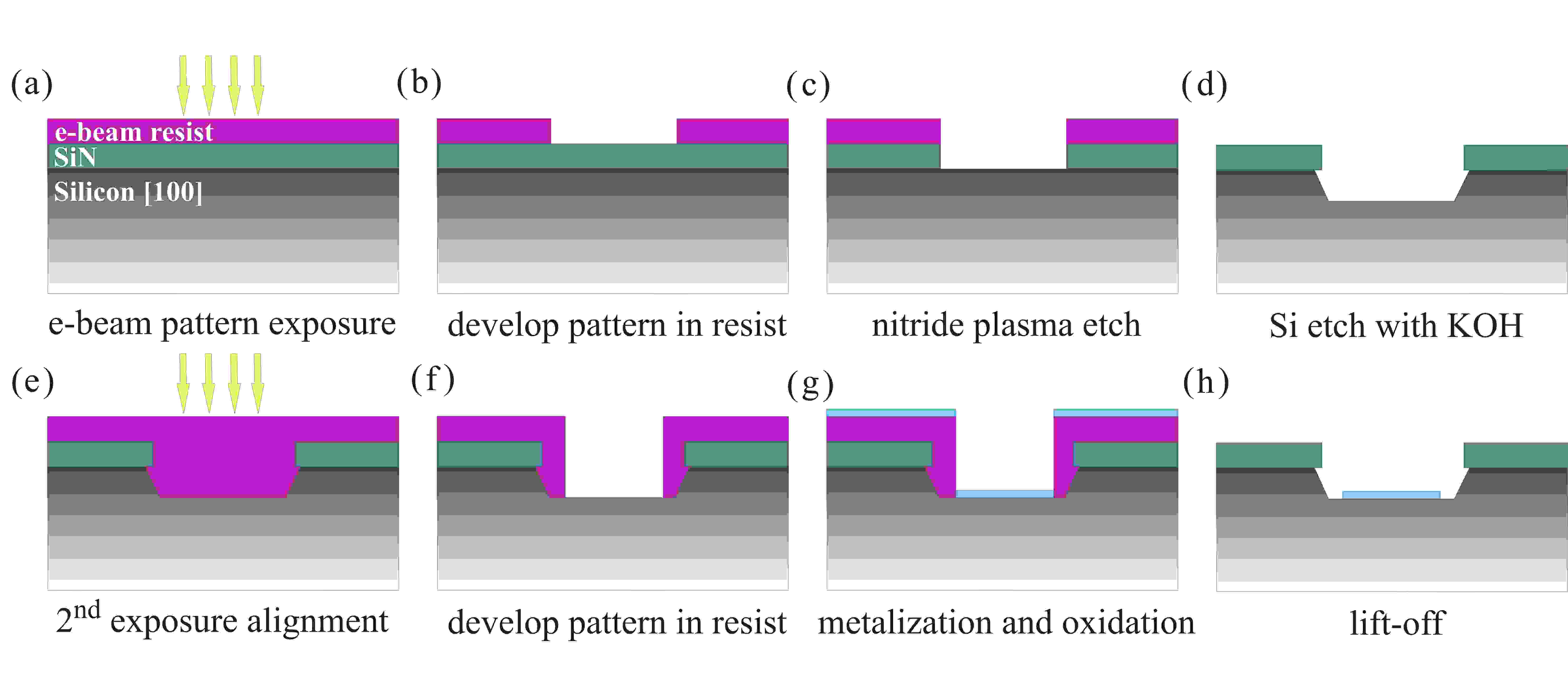}
\caption{Lithography process steps are shown. (a) Pattern is exposed in e-beam resist (PMMA), (b) resist pattern is developed, (c) SiN$_x$ is etched with CF$_ 4$ plasma and then (d) the Si substrate is etched anisotropically with KOH. (e) Second lithography is aligned with the previous pattern, (f) pattern is developed, (g) UHV evaporation of Ti and oxidation followed by (h) final lift-off.}
\label{fig:EtchProcessSteps}
\end{figure}
Lithographically assisted self-assembly was investigated with patterned surfaces made by a two-step electron-beam lithography technique, shown scematically in Fig. 2.  Samples were fabricated by first etching troughs into silicon substrates, using the overlay SiN$_x$ layer (750 nm) as an etch mask.  The nitride mask was defined by etching it reactively in CF$_4$ or CHF$_3$ plasma (for CHF$_3$, we typically had to etch for 20 min with parameters  min 100 W, 55 mTorr). The troughs with sidewalls angled at 54.74$\degree$ were produced by wet chemical etching in potassium hydroxide (KOH) at 90 $\celsius$. The depth of the troughs were varied from a few $\upmu$m up to nearly 100 $\upmu$m.  These troughs were then selectively coated with a hydrophilic TiO$_x$ layer by a second e-beam lithography step with litf-off aligned with the previous pattern, which produced TiO$_x$ coating only at the bottoms of the troughs. 

\section{Results}
\label{results}
\begin{figure}[h]
\centering
\includegraphics[width=1.0\textwidth]{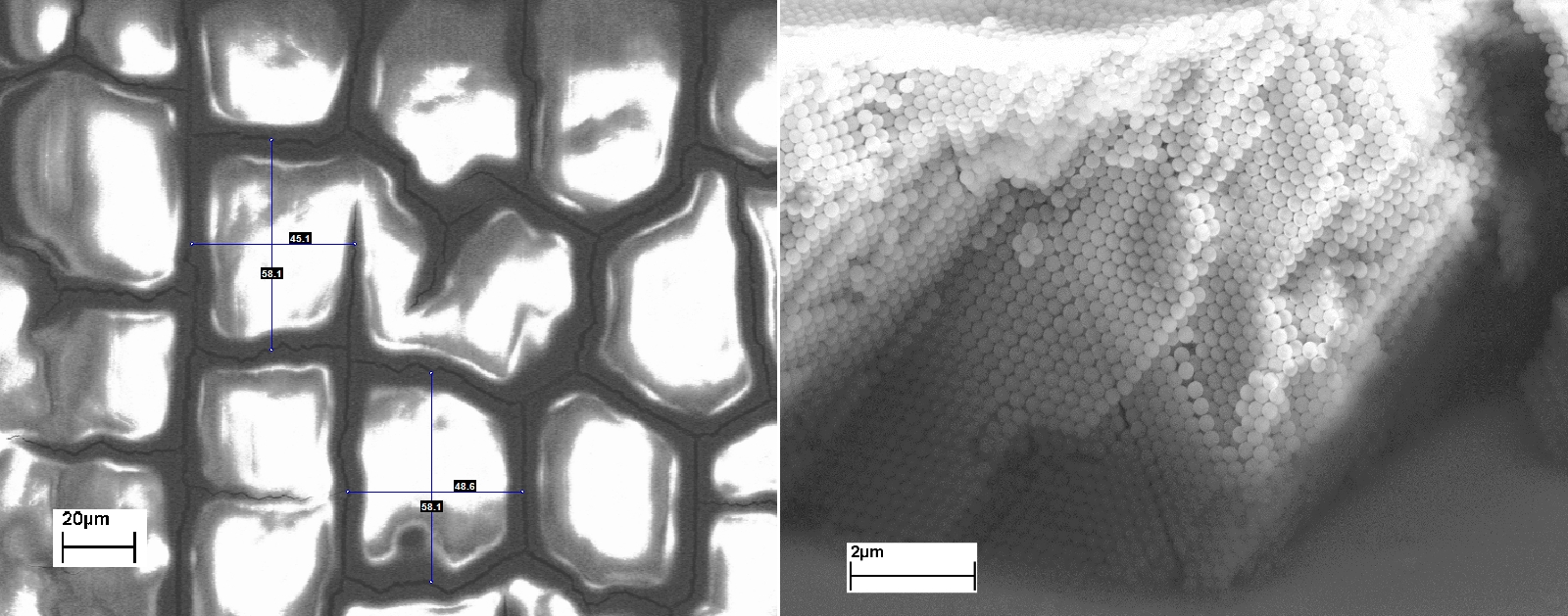}
\caption{An example of a SEM image used for the domain length measurements (left,) with a side-view close up of the multi-layer colloidal crystal structure (right). Values shown for measured lengths are in $\upmu$m.}
\label{fig:meas_inset}
\end{figure}

The fabricated crystals typically consisted of a set of domains, separated by cracks produced by the drying process. Domain size data were collected by taking several SEM images from each sample, and using the scale bar as a calibration length. A representative measurement image is shown in Figure \ref{fig:meas_inset}(left). All measurements take into account the direction of dipping, where measurements taken along the dipping direction were defined as vertical lengths, while those which are perpendicular were defined as horizontal. Also shown in Fig. 3 (right), is a side-view close up of a typical thick multi-layer colloidal crystal structure. Large data sets were collected from each sample in order to improve statistics and reduce the inherent roughness of the measurement method.  From the scatter plots of vertical vs. horizontal length, we always saw a distribution of domain sizes for all speeds and solution concentrations.

\begin{figure}[htp!]
\centering
\captionsetup[subfloat]{singlelinecheck=false}
	\begin{subfigure}[b]{0.5\textwidth}
	\caption{ }
	\includegraphics[width=1.0\textwidth]{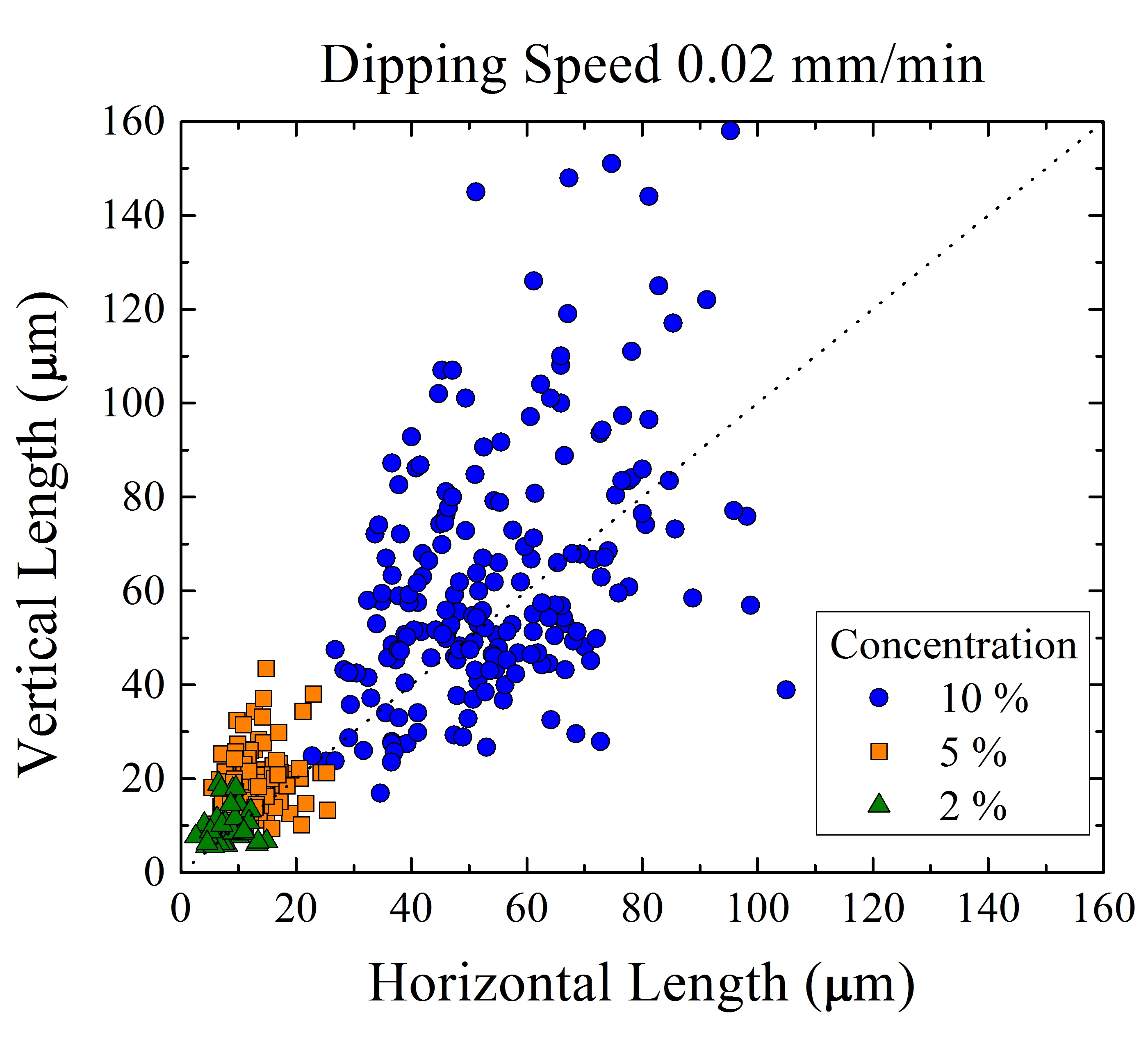}
	\label{fig:DipComp}
	\end{subfigure}%
~ 
	\begin{subfigure}[b]{0.5\textwidth}
	\caption{ }
	\includegraphics[width=1.0\textwidth]{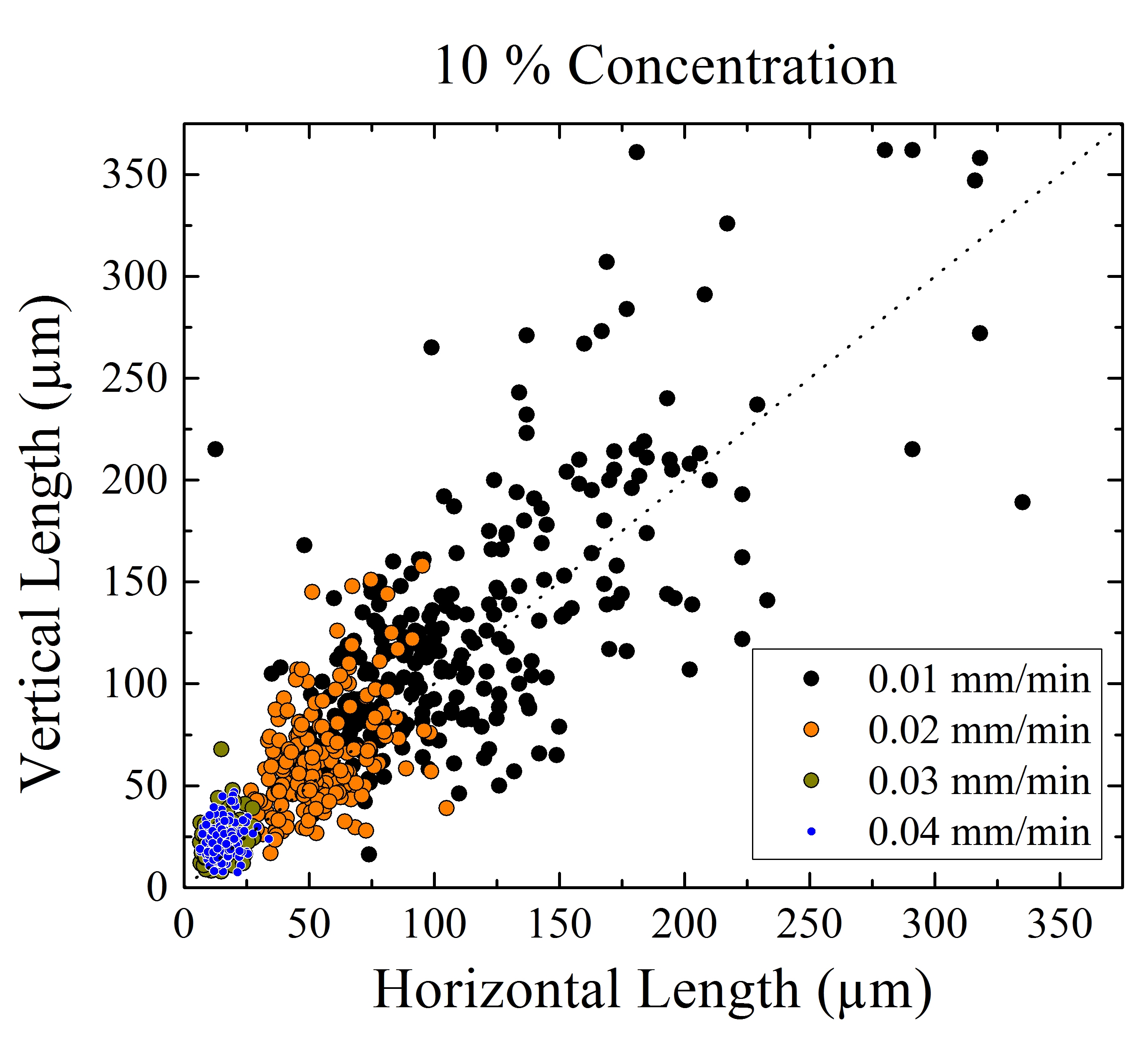}
	\label{fig:ConcComp}
	\end{subfigure}
\caption{(a) A comparison of domain sizes is shown for concentrations of 10\%, 5\% and 2\%, with all concentrations dipped at 0.02 mm/min, or (b)  for dipping speeds between 0.01 mm/min and 0.04 mm/min. The maximum domain size as well as the distribution increase with decreasing speed and increasing concentration.}
\label{fig:DomainSizes}
\end{figure}

\begin{figure}[hbp!]
\centering
\captionsetup[subfloat]{singlelinecheck=false}
	\begin{subfigure}[b]{0.5\textwidth}
	\caption{ }
	\includegraphics[width=1.0\textwidth]{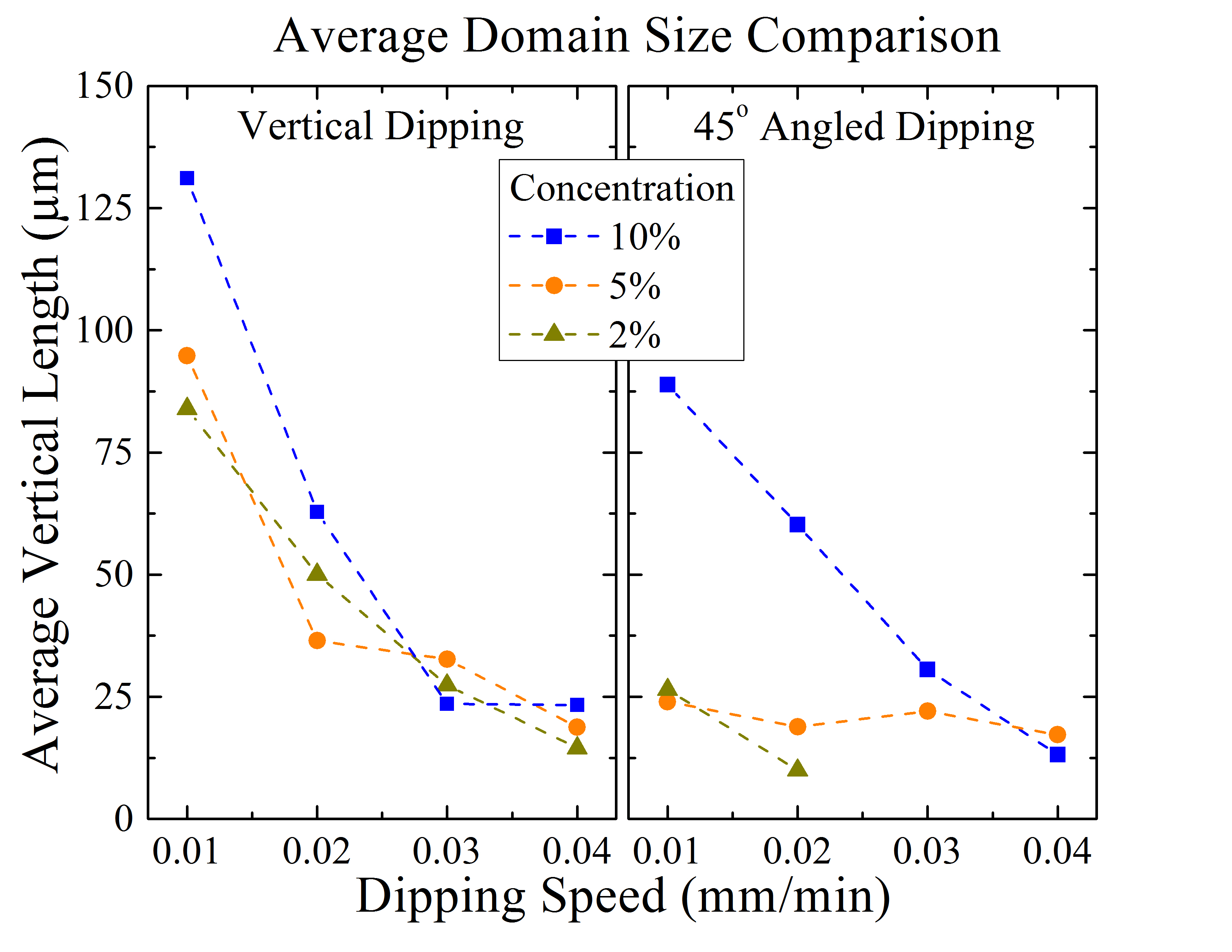}
	\label{fig:GraphMean}
	\end{subfigure}%
~ 
	\begin{subfigure}[b]{0.5\textwidth}
	\caption{ }
	\includegraphics[width=1.0\textwidth]{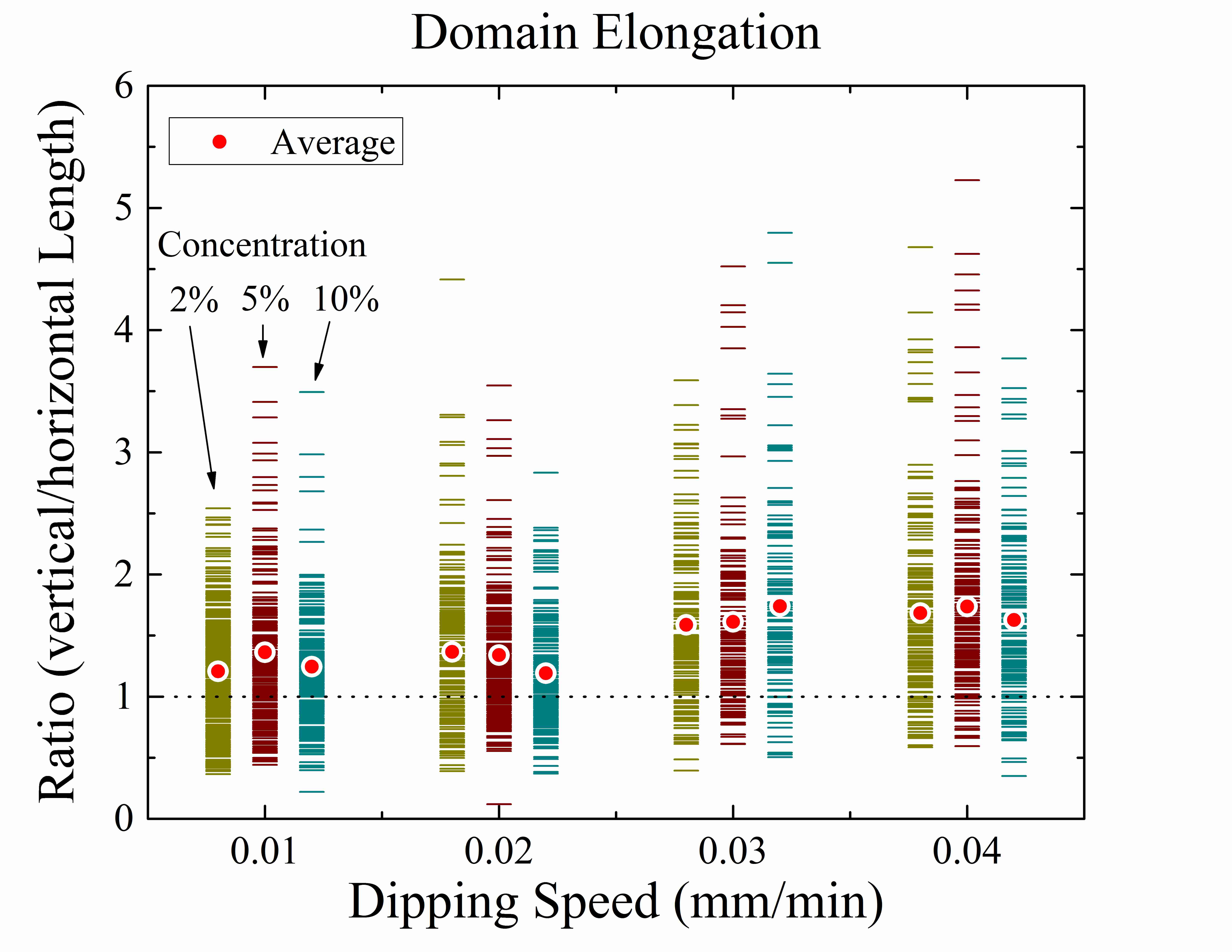}
	\label{fig:GraphRatio}
	\end{subfigure}
\caption{(a) Average vertical length of domains is plotted against dipping speed for vertical and angled dipping. Clear trends can be seen for dipping speed, concentration and angle. (b) Ratios of vertical to horizontal length of domains are plotted for vertical dipping. Vertical domain elongation and general spreading of domain sizes can be seen.}
\label{fig:Avg_Ratio}
\end{figure}

The plain, unpatterned samples dipped into solutions of 0.02\% and 0.2\% concentration did not produce any crystalline order even at the lowest dipping speeds and thus, no data is shown for these concentrations. For the higher concentrations crystalline domains always formed, and an example of the scatter plot of domain size measurement results for 2\%, 5\% and 10\% solutions dipped at 0.02 mm/min is shown in Figure \ref{fig:DipComp}.  From these, a clear trend can be seen with increasing concentration.  While there is a trend toward larger average domains with increasing concentration, the size distribution also increases.  This variability in domain size becomes quite large, as seen in the data for 10\% concentration, where the vertical domain size varies from a few tens of $\upmu$m up to approximately 150 $\upmu$m.  Similar trends are seen in the dipping speed comparison plots, with an example shown in  Figure \ref{fig:ConcComp} for the 10\% solution, at four dipping speeds.  At 0.01 mm/min, we see vertical lengths ranging from several tens of µm up to approximately 350 $\upmu$m.  Again, while the domain size grows with slower dipping speed, so does the size distribution.

Figure \ref{fig:GraphMean} shows the average vertical domain size for vertical and 45$\degree$ angled dipping, using three concentrations at four dipping speeds.  Angled dipping was found to produce smaller domains on average.  While the exact cause of this is unclear, it may be that changes in the shape of the meniscus can alter the evaporation rate at the self-assembly region.  This could lead to shorter effective self-assembly times in the angled dipping cases.  Though trends are clear, the wide distributions in domain sizes result in some scatter.  For example, at 0.02 and 0.03 mm/min vertical dipping speeds, we see that the average domain size does not correlate perfectly with concentration.  In the angled dipping, the 5\% concentration shows negligible change in average domain size with increasing speed.
While Figure \ref{fig:GraphMean} shows an apparent asymptotic approach to some minimum domain size, such a minimum is not supported by the data from the lower concentration solutions, where no domains were found.  Additionally, an indication of directionality in domain growth can be seen from ratios of the vertical to horizontal length.  This elongation effect can be seen from Figure \ref{fig:GraphRatio}, which shows these ratios for three concentrations at four dipping speeds.  (The positions of data for 2\% and 10\% concentrations are shifted in the plot for visual clarity).  The average ratio, shown in red for each data set, indicates an increased elongation along dipping direction with increasing dipping speed.  These data also show that the distribution of ratios widens with increasing dipping speed.
\begin{figure}[h]
\centering
\captionsetup[subfloat]{singlelinecheck=false}
	\begin{subfigure}[b]{0.3\textwidth}
	\caption{ }
	\includegraphics[width=1.0\textwidth]{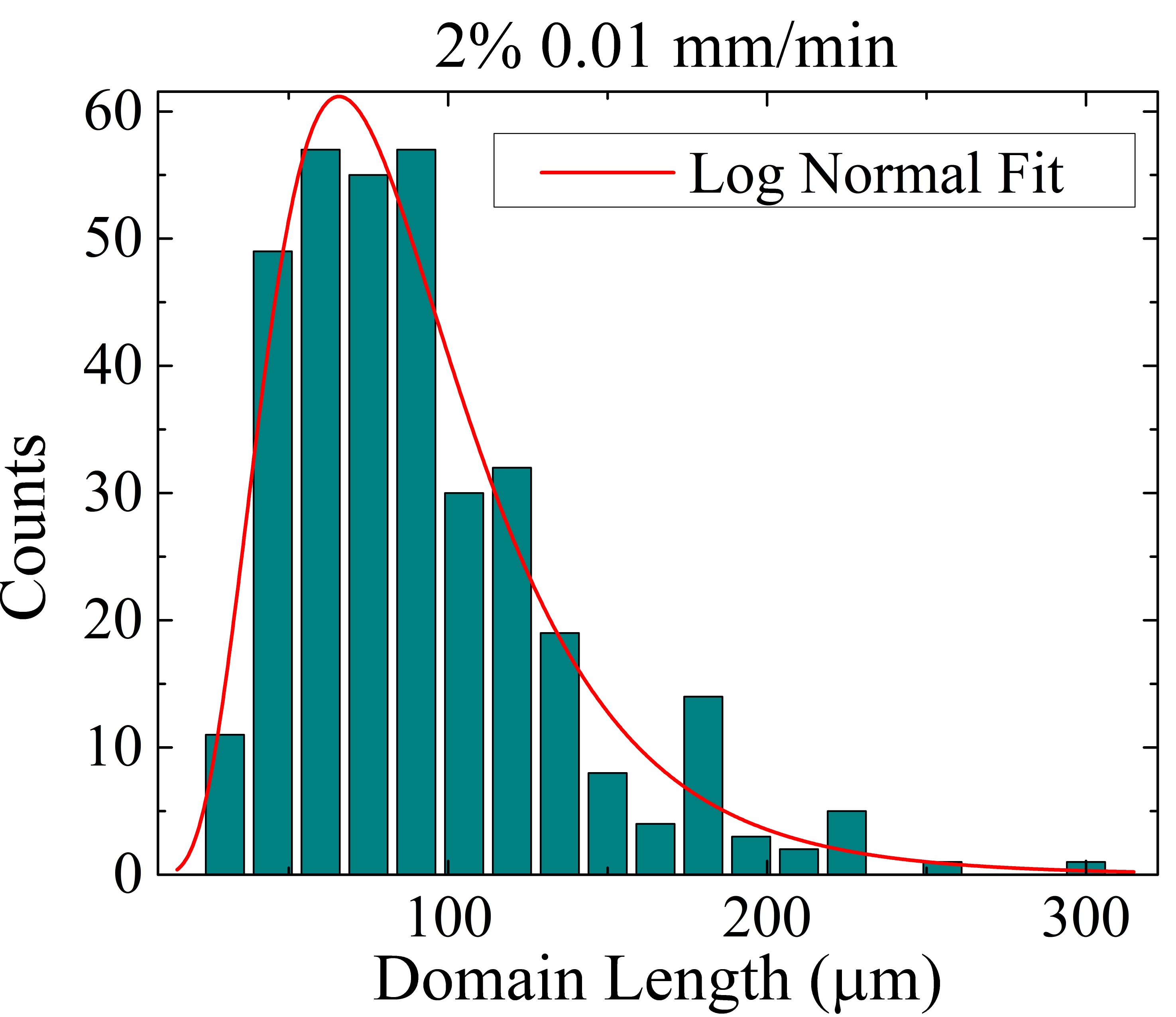}
	\label{fig:2pHist}
	\end{subfigure}%
~ 
	\begin{subfigure}[b]{0.3\textwidth}
	\caption{ }
	\includegraphics[width=1.0\textwidth]{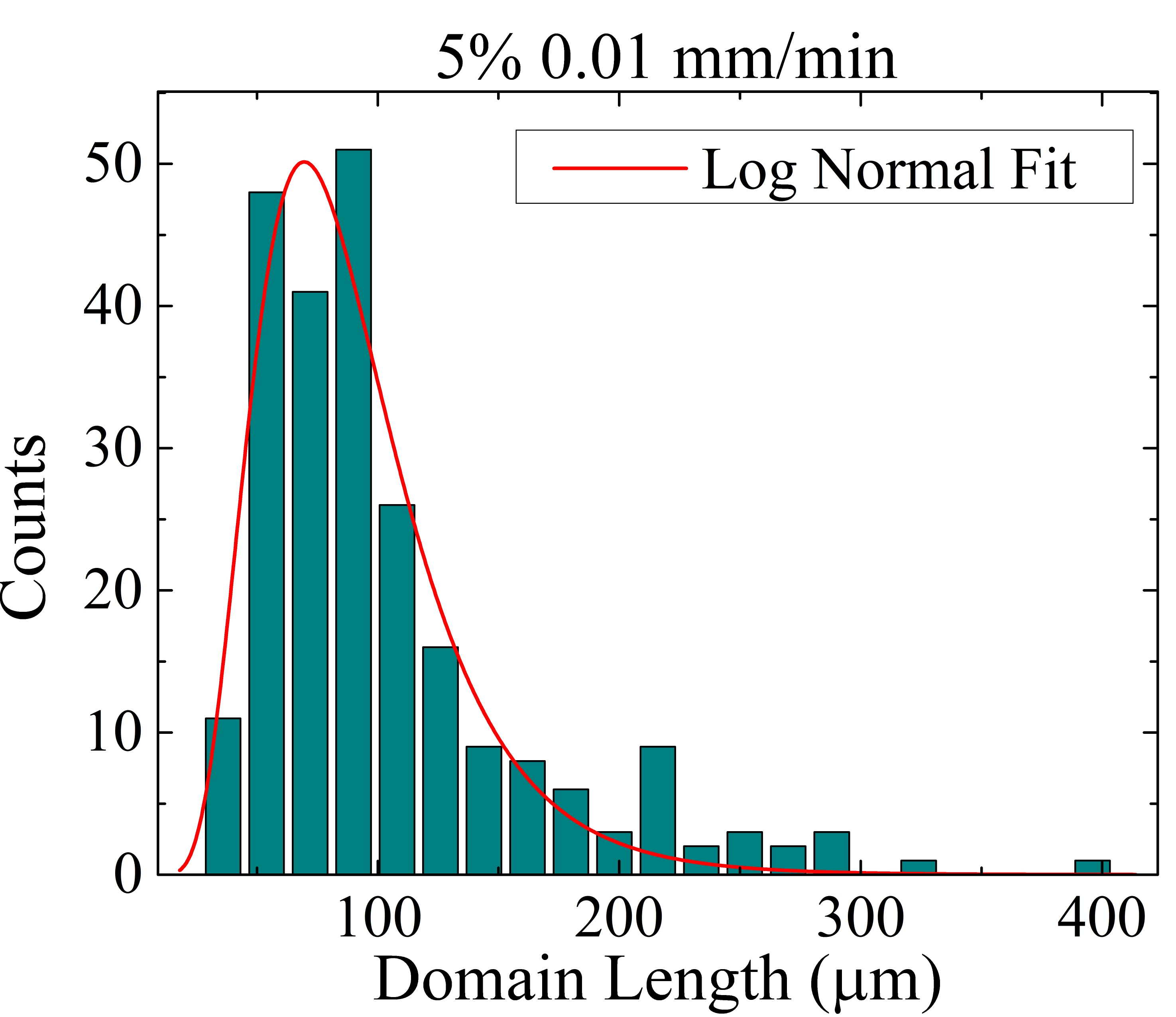}
	\label{fig:5pHist}
	\end{subfigure}
~ 
	\begin{subfigure}[b]{0.3\textwidth}
	\caption{ }
	\includegraphics[width=1.0\textwidth]{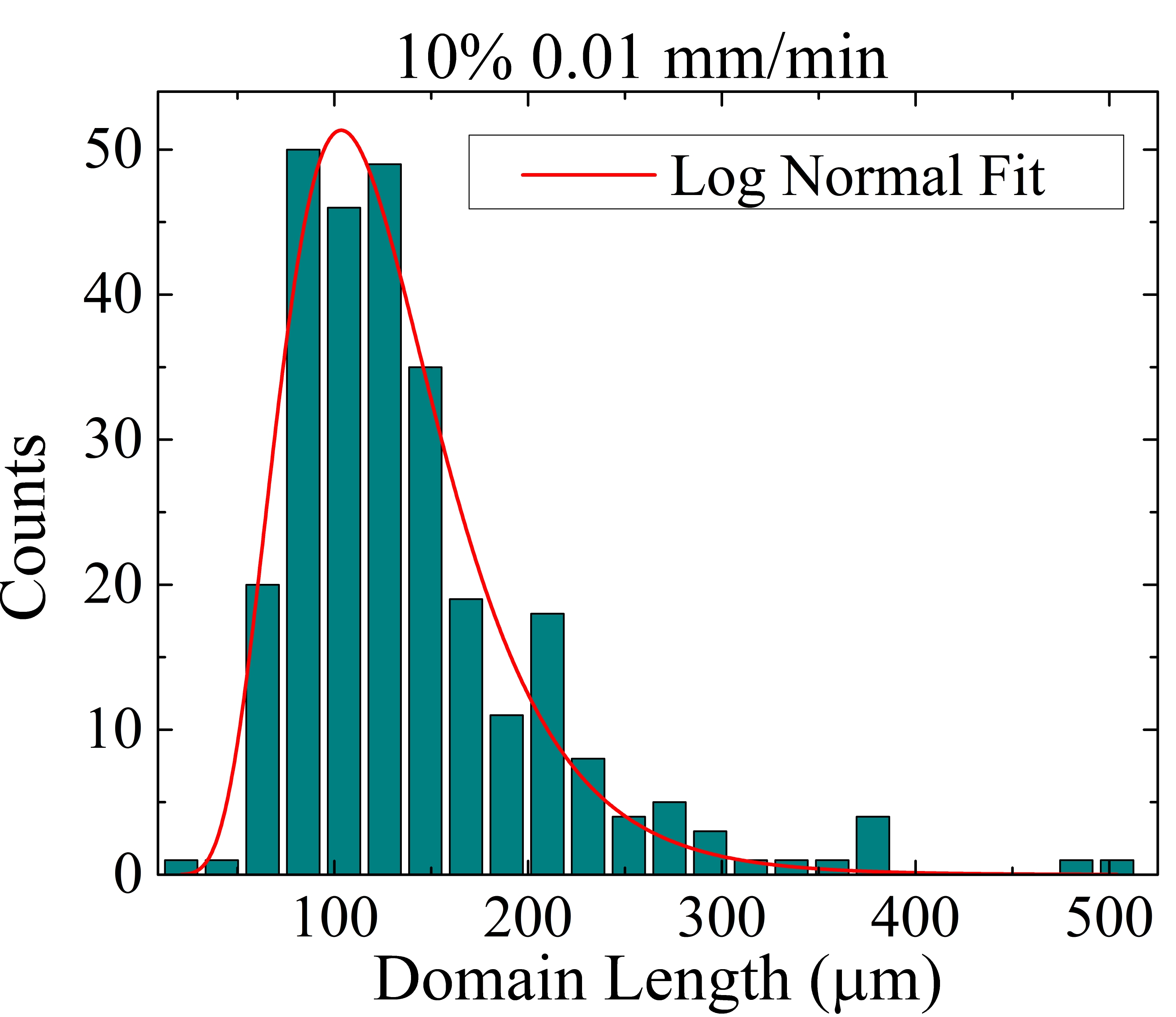}
	\label{fig:10pHist}
	\end{subfigure}
\caption{Histograms of vertical domain lengths are shown for (a) 2\%, (b) 5\% and (c) 10\% concentrations at a dipping speed of 0.01 mm/min. All three concentrations are characterized by a log-normal size distribution.}
\label{fig:Histograms}
\end{figure}

\begin{figure}[h]
\centering
\includegraphics[width=0.7\columnwidth]{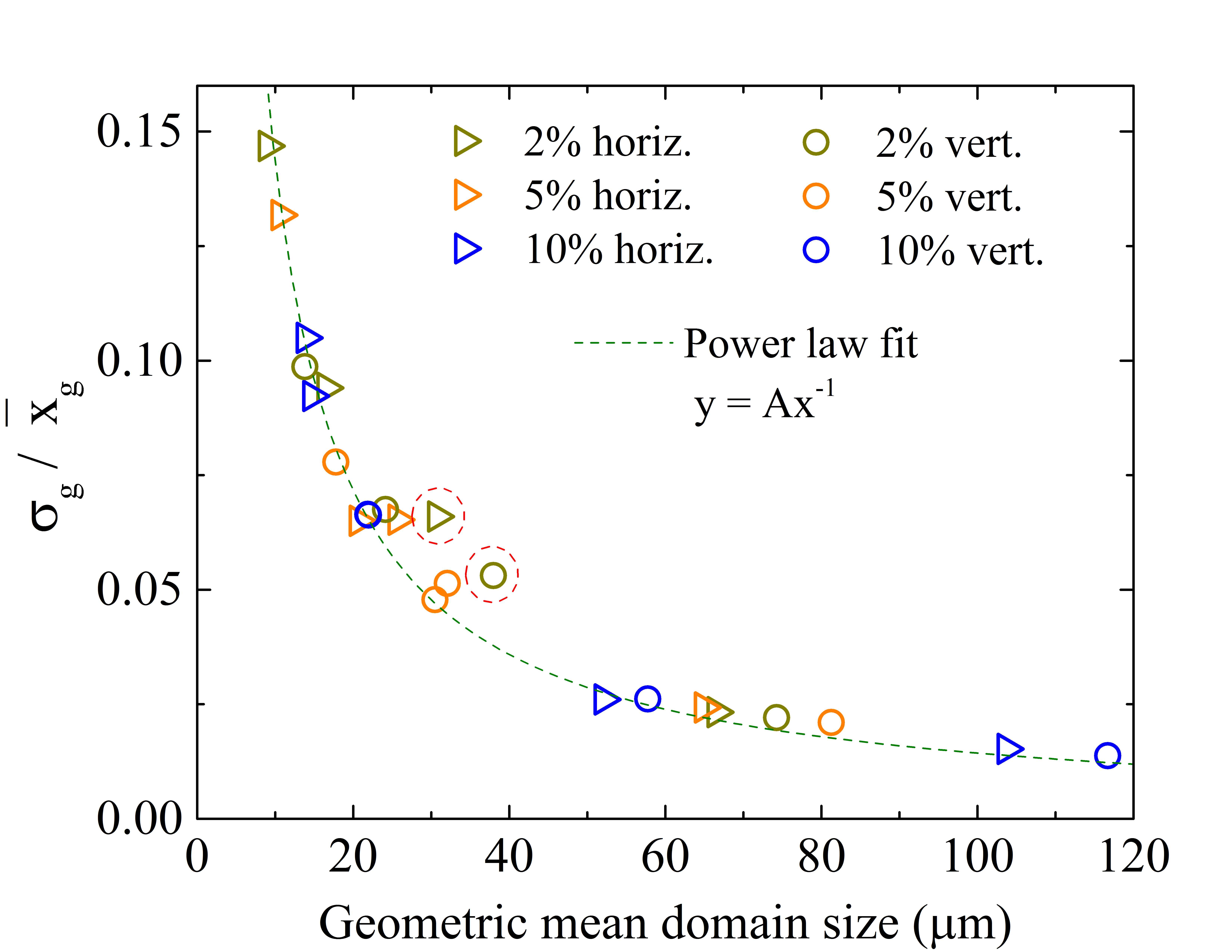}
\caption{Normalized geometric standard deviation $\sigma_g/\bar{x}_g$ data for all dipping speeds and concentrations vs. geometric mean. Two anomalously wide distributions (circled in red) can be seen to fall farther from the $1/\bar{x}$ fit (dashed curve).}
\label{fig:GeomColors}
\end{figure}

The domain size distributions for all dipping speeds and concentrations were also investigated, with example histograms of vertical domain lengths shown in Figure \ref{fig:Histograms} for 2\%, 5\% and 10\% concentrations at 0.01 mm/min dipping speed.  We found that all concentrations were approximately characterized by log-normal size distributions shown in red, especially at the lowest dipping speeds. Only at the highest dipping speeds do the distributions start to be less skewed and approach the normal distribution. Thus, most of the data are not well described by arithmetic mean and standard deviation.  Compiling, instead, the geometric means $\bar{x}_g=(\Pi_n x_n)^{1/n}$ and the geometric standard deviations $\sigma_g=\exp \left(\sqrt{\sum_n (\ln^2(x_n/\bar{x}_g)/n}\right)$ of these distributions, we found that all $\sigma_g$ values fell in the range of 1.32 and 1.71, with the exception of two anomalously wide distributions  (at 2\% concentration and 0.02 mm/min dipping speed, the horizontal and vertical domain size distributions had deviations $\sigma_g$ of 2.025 and 2.017, respectively). The normalized geometric standard deviations $\sigma_g/\bar{x}_g$ are very well approximated by a $1/\bar{x}_g$ trend for nearly all data points, as shown in Figure \ref{fig:GeomColors}, except for the two anomalous $\sigma_g/\bar{x}_g$ values seen (circled in red) in Figure \ref{fig:GeomColors} slightly farther from the fit curve. These two sets of data were likely the result of insufficient data sampling or poor substrate quality. 

The effects of etched lithographic features on self-assembly were studied by comparing results of samples dipped at various speeds
with troughs of varying size and depth.  For troughs of 10 $\upmu$m depth, there was hardly any change in domain sizes compared to areas outside the trough, while troughs of 30 $\upmu$m depth did not appear to fill completely.  All troughs of 20 $\upmu$m depth showed more consistent order in general than unpatterned areas.  The lateral shape of the trough was always a square, but with a side length varying between 50 -- 100 $\upmu$m in steps of 10 $\upmu$m, and for larger troughs from 150 to 300 $\upmu$m in steps of 50 $\upmu$m.  Troughs smaller than 70 $\upmu$m showed little to no difference in domain size when compared to the surrounding area.  In troughs larger than around 200 $\upmu$m, there were always more than one domain.
\begin{figure}[h]
\centering
\captionsetup[subfloat]{singlelinecheck=false}
	\begin{subfigure}[b]{0.5\textwidth}
	\caption{ }
	\includegraphics[width=1.0\textwidth]{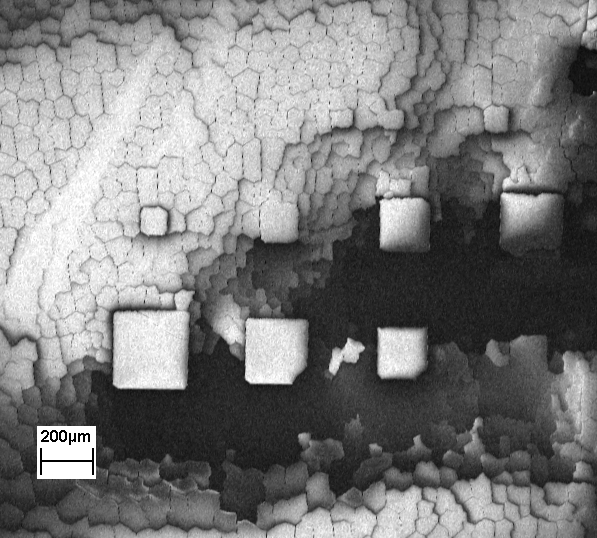}
	\label{fig:TroughsA}
	\end{subfigure}%
~ 
	\begin{subfigure}[b]{0.5\textwidth}
	\caption{ }
	\includegraphics[width=1.0\textwidth]{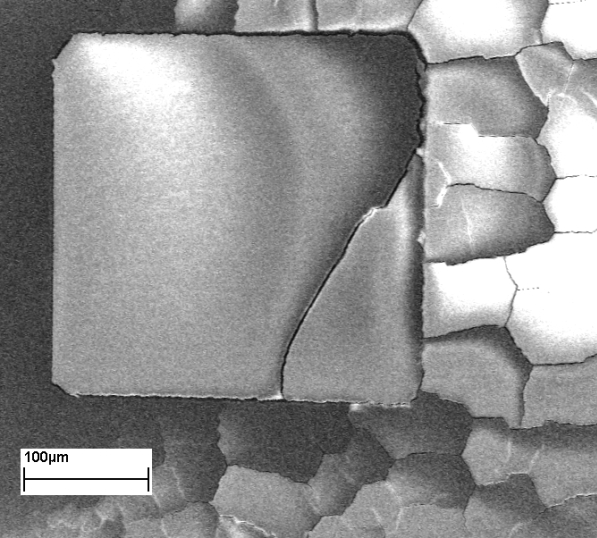}
	\label{fig:TroughsB}
	\end{subfigure}
\caption{(a) SEM image of a sample with a range of etch trough sizes, after dipping in 10 \% solution with speed 0.02 mm/min. Areas outside the troughs have domains ranging from 30 –- 70 $\upmu$m while the etched troughs show significantly larger domains. (b) Close up view of a 250 $\upmu$m by 250 $\upmu$m trough, seen with a continuous crystalline region several times larger than the surrounding domains.}
\label{fig:Troughs}
\end{figure}
Figure \ref{fig:TroughsA} shows clearly the increase of domain size within troughs compared to the unpatterned areas.  Domain size outside the trough is approximately 30 -- 70 $\upmu$m while inside, the crack-free domains can be many times larger.  The consistency of the continuous crystals in troughs indicates that this method can be used to locally suppress the domain size distribution completely.  Figure \ref{fig:TroughsB} shows a closer view of the largest trough and its surroundings.  While crack-free domains up to approximately 200 $\upmu$m size are reproducible, these regions are not necessarily single crystalline. All samples produced inside the troughs so far have polycrystalline order of crystallites with differing crystal orientations, and  the size of these single crystal domains $\sim$ 10 $\upmu$m is consistent throughout all the samples.

\section{Conclusions}
\label{conc}
Vertical deposition on TiO$_{x}$ coated substrates has been used to produce self-assembled colloidal crystals of polystyrene nano-spheres.  For unaltered surfaces, using a solution of 10\% PS spheres and a dipping speed of 0.01 mm/min, we achieved an average vertical domain size of approximately 130 $\upmu$m with some domains up to 350 $\upmu$m.  These unaltered substrates produce, however, large domain size variations, characterized by log-normal distributions, tending only slightly toward normal distributions at higher dipping speeds.  The relative variation in domain size increases with both dipping speed and concentration.
Nevertheless, the variability even at the lowest speeds is significant and poses a serious challenge for the fabrication of photonic or phononic crystals on largest scales.  
 
 On the other hand, lithographically assisted self-assembly was used to consistently produce continuous regions of multi-layer colloidal crystal as large as 200 $\upmu$m, effectively suppressing the size distributions found in unaltered substrates.  While these regions are free of cracks, they are still not fully single crystalline, and  the crystalline quality of these lithographically assisted multi-layer structures was comparable to that of the plain self-assembly samples.  Improving the crystalline quality of these regions remains a significant challenge for rapid production of large-scale structures.  Still, the method investigated in this paper is notably faster than most methods for large-scale colloidal crystal fabrication and thus, may be of interest to the photonic and phononic crystal communities.



\bibliographystyle{model1a-num-names}







\end{document}